\documentclass[twocolumn]{aastex631}
\usepackage{amsmath}

\newcommand\soutpars[1]{\let\helpcmd\sout\parhelp#1\par\relax\relax}

\defcitealias{giesers2019}{G19}
\defcitealias{Farrah2023a}{F23}

\begin{document}

\title{Constraints on the Cosmological Coupling of Black Holes from the Globular Cluster NGC 3201}

\author[0000-0003-4175-8881]{Carl L.~Rodriguez}
 \email{carl.rodriguez@unc.edu}
\affiliation{%
 Department of Physics and Astronomy,
    University of North Carolina at Chapel Hill,
    120 E. Cameron Ave, Chapel Hill, NC, 27599, USA
}%



\begin{abstract}

Globular clusters are among the oldest stellar populations in the Milky Way; consequently, they also host some of the oldest known stellar-mass black holes, providing insight into black hole formation and evolution in the early ($z\gtrsim 2$) Universe.  Recent observations of supermassive black holes in elliptical galaxies have been invoked to suggest the possibility of a cosmological coupling between astrophysical black holes and the surrounding expanding Universe, offering a mechanism for black holes to grow over cosmic time, and potentially explaining the origin of dark energy. In this paper, I show that the mass functions of the two radial velocity black hole candidates in NGC 3201 place strong constraints on the cosmologically-coupled growth of black holes.  In particular, the amount of coupling required to explain the origin of dark energy would either require both NGC 3201 black holes to be nearly face on (a configuration with probability of at most $10^{-4}$) or one of the BHs would need to have formed with a mass below that of the most massive neutron stars ($2.2M_{\odot}$).  This emphasizes that these and other detached black hole-star binaries can serve not only as laboratories for compact object and binary astrophysics, but as constraints on the long-term evolution of astrophysical black holes. 

\end{abstract}



\section{Introduction}

Globular clusters (GCs) are some of the oldest stellar systems in the Galaxy, and have often been invoked as probes of galaxy formation and assembly \citep{Ashman1998,Brodie2006,Kruijssen2020}, a potential source of cosmic re-ionization \citep{Ricotti2002,Boylan-Kolchin2018,Ma2021} and even as independent limits on the age of the Universe \citep{Gratton2003,Valcin2020}; as such, they also contain some of the oldest known stellar-mass black holes (BHs) in the Milky Way (MW).  Our understanding of BHs in GCs has shifted dramatically over the last 15 years, starting with the first extragalactic BH X-ray binary (BH-XRB) candidates \citep{Maccarone2007a,Barnard2011}, which were soon followed by the identification of several quiescent candidates in MW GCs \citep{Strader2012,Chomiuk2013,Miller-Jones2015a,Shishkovsky2018,Urquhart2020}.  More recently, \citet[][henceforth G19]{Giesers2018,giesers2019} identified several BH candidates in the GC NGC 3201 using the radial velocity (RV) measurements of stars about unseen companions.  These BHs are likely the oldest known stellar-mass BHs with well-characterized masses in the MW, and offer a probe into the formation and potential evolution of BHs from the early ($z\gtrsim2$) Universe.

Recently, \citet[][henceforth F23]{Farrah2023a} has used observations of elliptical galaxies to suggest that BHs may be coupled to the expansion of the Universe \citep[as proposed by][]{croker2019}.  This coupling can be in part motivated by the fact that the Kerr BH solution reduces to flat space as $r\rightarrow \infty$, and not the Friedman-Robertson-Walker (FRW) metric of standard cosmology.  Early studies of the behavior of non-singular BHs in an FRW background \citep{Nolan1993} {and with energized interiors \citep{Dymnikova1992}}, as well as more recent dynamical efforts \citep{Guariento2012}, have suggested that BHs (or BH-like objects) can exhibit the features necessary to couple to the FRW expansion, causing the interior stress-energy (and therefore the mass) of BHs to increase with the FRW scale factor.  \cite{croker2021} parameterized the growth of a BH at scale factor $a$ as 

\begin{equation}
M(a) = M(a_i)\left(\frac{a}{a_i}\right)^k
\end{equation}

\noindent where $a_i$ is the scale factor of the Universe when the BHs become coupled to the expansion (with $a\geq a_i$), and $k$ is a free parameter representing the strength of BH cosmological coupling (restricted to $-3\leq k \leq 3$).  For BHs observed in the local universe, this reduces to $M(z) = M_b(1+z)^k$, where $z$ is the redshift a BH forms with birth mass $M_b$.  Using observations of supermassive BH (SMBH) growth in elliptical galaxies in both the local and early ($z\lesssim2.5$) Universe, \cite{Farrah2023} noted a significant offset between the stellar mass growth of galaxies and their SMBHs at high redshift.  One potential source of this offset, as suggested by \citetalias{Farrah2023a}, could be the aforementioned cosmological coupling, which would cause SMBHs to grow substantially from $z\lesssim7$ to the present day.

\citetalias{Farrah2023a} explored this mass discrepancy in the context of this cosmological coupling; using the SMBH data from \citet{Farrah2023}, they found that the early growth of these elliptical galaxy SMBHs could be explained by a coupling strength of $k=3.11^{+1.19}_{-1.33}$.  They also argue that, if all BHs (and particularly stellar-mass BHs) experienced this $k\sim3$ coupling to the FRW scale factor, they would contribute to the accelerating expansion of the Universe as a source of dark energy, potentially explaining the measured value of $\Omega_\Lambda=0.68$ \citep{PlanckCollaboration2020}.  In this paper, I argue that a $k\sim 3$ coupling is incompatible with the measured masses and ages of the 2 confirmed BH candidates in NGC 3201, and would require at least one of the BHs to have been born with a mass less than $2.2M_{\odot}$, the lower-bound on the mass of the most massive non-rotating neutron stars (NSs).  For the present-day masses to be large enough to accommodate this amount of BH growth ($k\geq 2.5$), both BH-star binaries would need to be aligned nearly face on, a configuration with a less than $10^{-4}$ chance of occurring randomly.   In Section II, I describe the ages and measured masses of NGC 3201, while in Section III, I convert these measurements into limits on the BH coupling constant, $k$.  Throughout this paper, I assume a flat FRW cosmology from \cite{PlanckCollaboration2020} with $H_0 = 67.5\rm{~km}\rm{~s}^{-1} \rm{Mpc}^{-1}$ and $\Omega_m=0.315$.

\section{The Age and Masses of NGC 3201's BHs}\label{sec:level1}

Using the Multi Unit Spectroscopic Explorer (MUSE) integral-field spectrograph \citep{Kamann2018}, \citetalias{giesers2019} identified three candidate BHs in NGC 3201 with $M \sin (i)$ masses of $4.53\pm 0.21M_{\odot}$ (ACS ID \#12560, henceforth BH1), $7.68\pm0.5M_{\odot}$ (ACS ID \#21859, BH2), and $4.4\pm2.8M_{\odot}$ (ACS ID \#5132, BH3).  Of these, BH1 and BH2 are considered to be dynamically-confirmed BHs.  The RV curve for BH3 admits two solutions for the mass of the unseen companion, with the second solution having a minimum mass of $1.10\pm0.20M_{\odot}$ \citepalias[at 18\% probability; see][Section 7]{giesers2019}.    

While the $M \sin (i)$ masses of the BHs represent the minimum allowed mass {(i.e.~the mass of the system assuming an orbital inclination of $90^{\circ}$, where the RV we measure is the full velocity of the system)}, what we are interested in is the probability that the minimum of the BH masses in NGC 3201 is above some particular value, in order to place limits on how much BH growth may have occurred since their formation. Since the inclination of the binaries are assumed to be isotropic, we can write the geometric probability that the mass of any binary is above some mass $M$ as 

\begin{equation}
P(m > M | M') = 
\begin{cases}
1-\cos(\arcsin(M'/m)) ,& \text{if } m > M'\\
    1,              & \text{otherwise}\\
\end{cases}\nonumber
\end{equation}
\\
\noindent where $M'$ is the measured minimum mass.  Since we are only interested in the minimum mass of \emph{either} BH, we can write down the probability that the smallest BH mass in NGC 3201 (using only the two reliable BH candidates) is greater than $M$ simply as

\begin{align}
P\left(\min(M_{\rm BH}) > M | M'_1, M'_2\right) =&\nonumber \\ P(m > M | M'_1) \times& P(m > M | M'_2)
\end{align}

\noindent where $M'_1$ and  $M'_2$ are the minimum masses of BH1 and BH2, respectively.

\begin{figure*}[t!]
\includegraphics{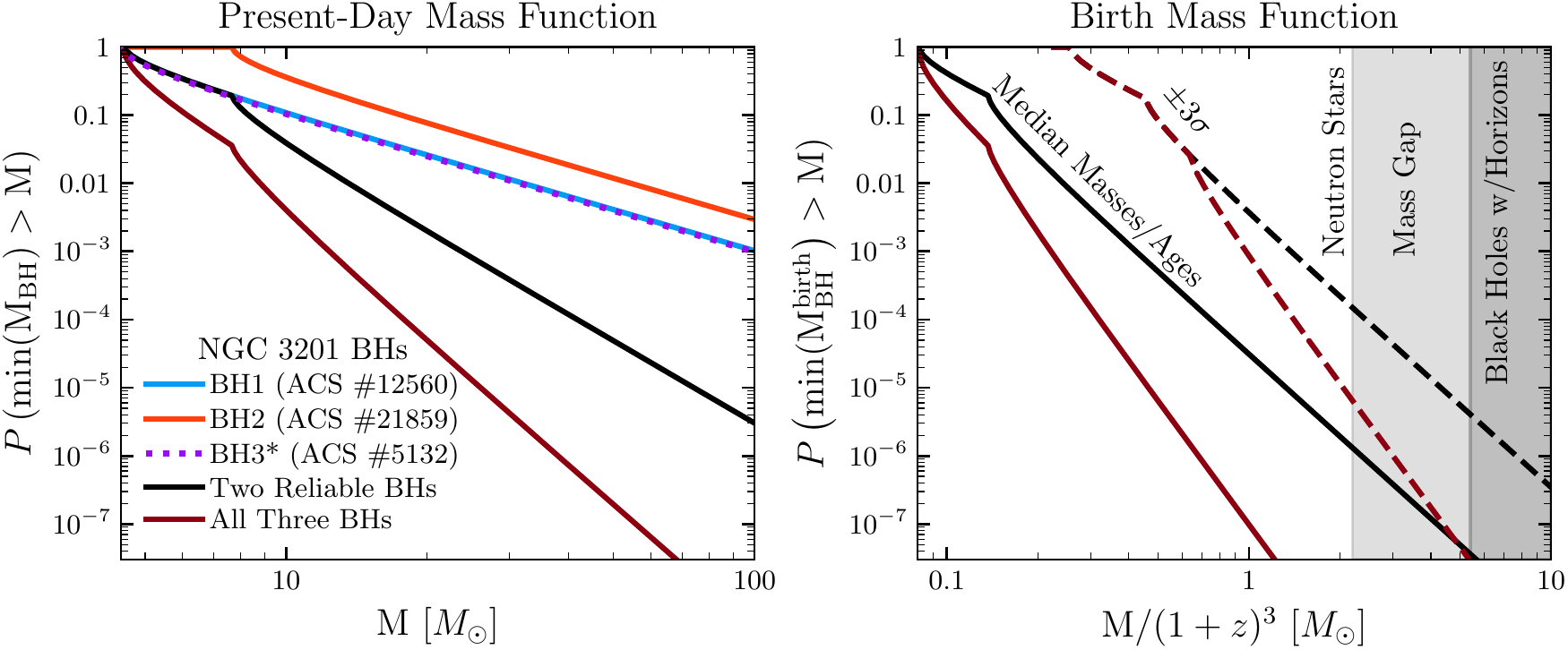}
\caption{The present-day and possible birth mass function of the BHs in NGC 3201.  On the \textbf{left}, I show the probability that the two reliable (BH1 and BH2) and third possible (BH3) BHs from \citetalias{giesers2019} have masses above $M$ in red, blue, and dotted purple respectively.  I also show the combined probabilities that both (or all three) BHs have mass above $M$ in black and burgundy, respectively.  On the \textbf{right}, I show what these masses would have been at birth to accommodate the growth required of a population of $k=3$ BHs.  Even allowing for $3 \sigma$ uncertainties in both the RV masses and the cluster age measurements, either both RV BHs would need to be nearly face on (a configuration with a probability of $<10^{-4}$) or one of the BHs must have been born with a mass below that of the most massive NS.}
\label{fig:1}
\end{figure*}

In the left panel of Figure \ref{fig:1}, I show the present-day mass function of the BHs in NGC 3201, as well as the combined minimum mass function considering both reliable BHs (BH1 and BH2) and all three BH candidates (BH1, BH2, and BH3).  Despite the lack of specific limits on the maximum mass, we can place good limits on how large the minimum BH mass can be.  Combining only the two reliable BH candidates, there is only a 10\% change the smallest BH mass is above $8.3M_{\odot}$, and a 0.1\% it is above $24M_{\odot}$.  Including BH3 (by multiplying Equation 2 by $P(m > M | M'_3)$, where $M'_3$ is the minimum mass of BH3) reduces these to $6.1M_{\odot}$ and $12M_{\odot}$, respectively.  Note that these mass functions do not yet include any uncertainties from the RV measurements themselves.

To use these mass functions to constrain the cosmological growth of BHs, we also need to know the redshift at which the GC and its BHs formed.  Measuring the absolute ages of GCs is a longstanding problem in observational astronomy, as it requires resolving the stellar population down to at least (and preferably much deeper than) the main-sequence turn off (MSTO).  Much of the recent work on MW GC ages has been enabled by the deep Hubble Space Telescope (HST) ACS survey of galactic GCs, which has dramatically improved their age estimates over the past 30 years. Typically, specific features or populations in the cluster, such as the distance between the MSTO and the subgiant branch \citep{dotter2010} or the main-sequence knee \citep[MSK][]{bono2010}, are isolated so that they can be compared to the predictions from stellar isochrones to determine the absolute age and distance of a cluster (though see \cite{rozyczka2022}, which instead fits isochrones to eclipsing binaries whose masses and distances are measured \citep{Paczynski1997}, or \cite{cabrera-ziri2022}, which fits isochrones to the integrated spectra of NGC 3201 while controlling for the error introduced by hot horizontal branch stars).  

In Table \ref{tab:ages}, I list age estimates for NGC 3201 from several studies using a variety of methods over the past 15 years.  Most studies rely on fitting specific CMD features to stellar isochones, and while there does exist disagreement between the various studies, the various estimates are all centered within 1.5 Gyr of 11.5 Gyr.  We take $11.5\pm0.5$ Gyr from \cite{rozyczka2022} as our standard age, as their estimate combines information from isochrone fits to the CMD with separate distance and age measurements using 3 eclipsing binaries near the MSTO.  This yields a formation redshift of $z=2.8$ for NGC 3201.  Note that I am ignoring the time it takes for a massive star to collapse to $4.53M_{\odot}$ BH, since that time (12.4 Myr, using the standard assumptions from the \texttt{COSMIC} population synthesis code, \citealt{Breivik2020}, and a stellar metallicity of $[\rm{Fe}/{H}] = -1.58$ from \citealt{Aguilera-Gomez2022}) is substantially less than the uncertainty on the cluster age.

\begin{table}
\caption {Recent Age Estimates for NGC 3201.  The isochrones used are indicated with $\dag$ for the Dartmouth Stellar Evolution Database \citep[DSED,][]{dotter2008}, $\ddag$ the Bag of Stellar Tracks and Isochrones \citep[BaSTI,][]{Pietrinferni2004}, $\mathsection$ for the Victoria-Regina Isochrone Database \citep[VR,][]{VandenBerg2014}, * for the MESA Isochrone and Stellar Tracks \citep[MIST,][]{Choi2016}, and $\mathparagraph$ for the models from \cite{VandenBerg2012}.}%
\label{tab:ages} 
\begin{ruledtabular}
\begin{tabular}{lcr}
\textrm{Study [Ref]}&
\multicolumn{1}{r}{\textrm{Age [Gyr]~}}&
\textrm{Method}\\
\colrule
\cite{cabrera-ziri2022} & $12.9\pm0.6$ & Int.~Spectra$^*$\\
\cite{rozyczka2022}\footnote{The authors combine these two methods into a single estimated age of $11.5\pm0.5$ Gyr, which I adopt as the cluster age.}  & $12.5\pm1.0$ & CMD$^\dag$\\ 
--- & $11.0\pm0.5$ & Ecl. Binaries$^\dag$\\
\cite{monty2018}\footnote{First row for each isochrone indicates fits to $K_s$ vs.~F606W-$K_s$ CMDs, while second row fits F336W vs.~F336W-$K_s$ CMDs.}  & $12.0\pm0.7$ & CMD$^\dag$\\ 
---  & $12.4\pm0.8$ & CMD$^\dag$\\ 
---  & $11.2\pm0.9$ & CMD$^\mathsection$\\ 
---  & $10.7\pm1.3$ & CMD$^\mathsection$\\ 
---  & $12.4\pm0.6$ & CMD$^\ddag$\\ 
---  & $11.2\pm2.0$ & CMD$^\ddag$\\ 
\cite{wagner-kaiser2017}  & $12.8^{+0.3}_{-0.2}$ & CMD$^\dag$\\
\cite{vandenberg2013} & $11.5\pm0.4$ & CMD$^\mathparagraph$\\ 
\cite{dotter2010} & $12.0\pm0.75$ & CMD$^\dag$\\
\cite{bono2010}\footnote{Rows correspond to estimates based on \textit{K, I,} and \textit{WK} bands respectively. Only the ages calculated from the MSTO-MSK differences in the CMD are shown} & $10.7\pm0.6$ & CMD$^\mathparagraph$\\ 
--- & $11.6\pm0.6$ & ---\\
--- & $10.8\pm0.7$ & ---\\

\end{tabular}
\end{ruledtabular}
\end{table}

\section{Constraints on Cosmological Growth of Stellar-Mass BHs}

With an assumed formation redshift for NGC 3201 and the minimum mass function of its BHs, we can now place good constraints on how much those BHs may have grown with the expansion of the Universe.  The only thing we still need is an assumed minimum mass for stellar-mass BHs.  From both BH-XRB observations and gravitational-wave (GW) detections, there exists evidence of a mass gap between the largest NSs and the lowest-mass BHs \citep{Ozel2010,Farr2011,TheLIGOScientificCollaboration2021}.  Recent studies of the 4 NS-BH mergers observed by LIGO/Virgo have suggested that---if the lower mass gap exists---the minimum BH mass is $5.4^{+0.7}_{-1.0}M_{\odot}$ \citep{Ye2022}, a result that agrees well with the same value inferred from the BH-XRB population.  However, it should be noted that a full analysis of the LIGO/Virgo population (including all NSs and BHs) reduces the evidence for a lower mass gap \citep{TheLIGOScientificCollaboration2021,Farah2022}, and that theoretical modeling of supernova does not require the existence of the gap \citep{Fryer2001,Fryer2012}.  Since we are interested in the upper limit on BH coupling, the most conservative choice is to assume the mass gap does not exist, and set the minimum BH birth mass to $2.2M_{\odot}$, the inferred maximum mass for a non-rotating NS \citep[e.g.,][]{Legred2021}.  This is obviously an extreme assumption, but it places the best possible constraints on the amount of BH growth that could have occurred since $z=2.8$ {(also note that this assumption is \emph{more} conservative than the $2.7M_{\odot}$ minimum mass assumed by \citetalias{Farrah2023a})}

In the right panel of Figure \ref{fig:1}, I show the birth mass function for the minimum BH mass in NGC 3201, assuming a $k=3$ coupling between the scale factor and the BH masses, and illustrate the core argument of this paper: to accommodate the large (factor of $\approx55$) amount of mass growth needed for cosmic coupling to explain the existence of dark energy, either one of the BHs in NGC 3201 would need to have been born with a mass below the most massive NS (where BH formation is presumably excluded{, though see the discussion in \S \ref{theoriststuff}}), or both BH1 and BH2 must be aligned nearly face on to Earth, such that both masses are above $122M_{\odot}$.    Given our knowledge of the $M \sin(i)$ masses, such a configuration has a less than a $10^{-6}$ probability of occurring.  Including BH3 substantially strengthens these bounds,  and would require all three systems to be nearly face on with probability less than $10^{-9}$.   Even allowing for generous ($+3\sigma$) error on the measured minimum mass of the two reliable BHs from \citetalias{giesers2019} and an equally generous ($-3\sigma$) error on the age from \cite{rozyczka2022}, the probability that the least massive of BH1 and BH2 was born above $2.2M_{\odot}$ is less than $2\times10^{-4}$ (and less than $6\times10^{-6}$ for the 3 BH configuration).

Of course, this argument can be reversed to ask what kind of limits on $k$ can the mass function of the 2 NGC 3201 BHs provide?  In Figure 2, I show the upper limits on $k$, assuming Gaussian uncertainties on the two BH masses (neglecting BH3), and implicitly assuming a uniform prior on the BH masses (apart from the minimum BH mass, which I set to either $2.2M_{\odot}$ or $5.4M_{\odot}$, depending on whether the mass gap is assumed to exist).  With no mass gap, and BHs allowed to form as light as the most-massive NS, we can already exclude a $k\geq 2.5$ with a probability of $4\times10^{-5}$, while a $k\geq2$ is excluded at $5\times10^{-4}$.  On the other hand, if the mass gap does exist, and the minimum BH mass is closer to $5.4M_{\odot}$ (already above the minimum mass of BH1), these limits become extreme, excluding $k\geq 2.5$ with a probability of $3\times10^{-6}$.  In both cases, these limits are sufficiently stringent that they can not only exclude BHs as a potential source of dark energy, but can potentially place limits on other classes of theories that predict coupling between the non-singular BH theories and an expanding FRW cosmology. 

\begin{figure}[t]
\includegraphics{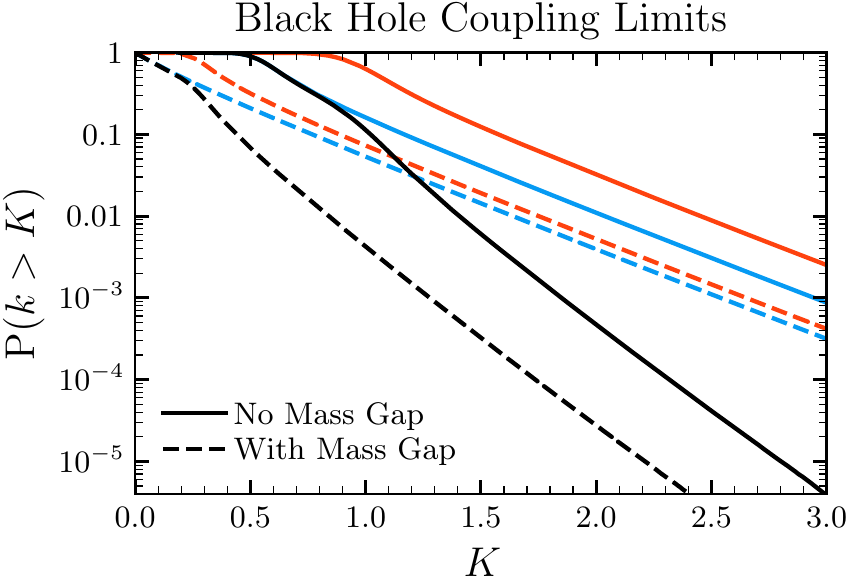}
\caption{Limits on the coupling constant between BHs and cosmological expansion.  I consider minimum BH masses of $2.2M_{\odot}$ (assuming no mass gap between BHs and NS) and $5.4M_{\odot}$, \citep[if the mass gap exists, from][]{Ye2022}.  Only BH1 and BH2 are considered here, using the same color scheme as Figure 1.}
\end{figure}

\section{Implications of Cosmological Coupling}

\subsection{Could horizonless BHs form with subsolar masses?}
\label{theoriststuff}

A fundamental assumption of this work is that there can be no BHs formed from stellar collapse with birth masses below that of either $2.2M_{\odot}$ or $5.4M_{\odot}$.  The former (and most conservative) limit arises largely from the equation of state inferred by NICER and XMM-Newton observations of PSR J0740+6620, a NS in a binary system whose Shapiro time delay allows for a well-constrained mass measurement \citep[$2.08\pm0.07M_{\odot}$,][]{Fonseca2021}.  The BH lower-mass limit of $5.4M_{\odot}$ from \cite{Ye2022} assumes that LIGO/Virgo's GW observations are representative of the BH birth masses (modulo later stellar accretion).  If BHs were to grow via cosmological coupling, then the mass gap would likely not exist.  However, an analysis of the LIGO/Virgo binary BH mergers presented in \cite{croker2021} using a nearly identical population synthesis approach to that used in \citetalias{Farrah2023a} (assuming no mass gap and a minimum BH mass of $2.5M_{\odot}$) found support for only moderate BH coupling ($k\approx0.5$), in tension with the \citetalias{Farrah2023a} results.  As shown in Figure 2, such a weak coupling is also consistent with BH1 and BH2 \citep[which is hardly surprising, given that GCs may contribute a significant fraction of LIGO/Virgo's BH mergers, e.g., ][]{Zevin2021,Wong2021,Rodriguez2021}.  However, this weak $k\approx0.5$ coupling would be insufficient to provide a stellar-mass BH explanation of $\Omega_\Lambda$.

Of course, one could imagine extreme scenarios where cosmologically-coupled BHs (which are not Kerr BHs, {but rather horizon-less BH-like objects in GR}) form with masses of $\sim 0.1 M_{\odot}$, perhaps in the centers of imploding massive stars, and somehow manage to avoid accreting more infalling stellar material.  This may be conceivable given the {difficulty of accreting new material onto} a horizonless BH, {which would require some mechanism for converting ordinary matter into whatever energized material constitutes the BH interior}.\footnote{{I note that such a difficulty would also have radical implications for the assumed accretion rates of BHs in both X-ray binaries and Active Galactic Nuclei.}}

{Of course, there still exists the possibility that both NGC 3201's BH and LIGO/Virgo's BHs may have formed with subsolar masses, and cosmolgically grown with redshift to their current observed values.  This is indeed a possibility for the GW sources, since it may take several Gyrs for GWs to drive binary BHs to merge after their formation.  However, such low birth masses are in conflict with BHs in high-mass X-ray binaries (HMXBs), where a compact object accretes matter from the stellar winds of a young and massive ($\gtrsim 5M_{\odot}$) companion.  Of the many known HMXBs \citep[$ > 100$ in the Milky Way alone; see the recent catalogs of][]{Fortin2023,Neumann2023}, three are known to host dynamically-confirmed BHs: Cygnus X-1, with a BH mass of $21.2\pm2.2 M_{\odot}$ and an age of $\sim 4$ Myr \citep{Miller-Jones2021}, LMC X-1, with a BH mass of $10.91\pm1.41M_{\odot}$ and an age of $\sim 5$ Myr \citep{Orosz2009}, and M33 X-7, with a BH mass of $11.4M_{\odot}$ and an age of $\sim 5.8$ Myr \citep{Ramachandran2022}.  Because their accretion is fed from the winds of the companion, these BHs likely have masses close to their natal values.  As an example, the companion to Cygnus X-1 loses approximately $2-5\times10^{-6}M_{\odot}/\rm{yr}$ \citep{Gies2003,Vrtilek2008}, of which $\sim 0.4\%$ is accreted onto the BH companion \citep{Hirai2021}.  Even if the BH were able to accrete at that rate for 4 Myr, it would only have grown by $\sim 0.1M_{\odot}$.  As noted by \cite{Miller-Jones2021}, the actual accretion time is likely far shorter ($0.02-0.04$ Myr) given the age of the jet-inflated optical nebula surrounding Cygnus X-1 \citep{Russell2007}.}

{The only way for both the old BHs in NGC 3201 and the young BHs in Cygnus X-1 (and other HMXBs) to be simultaneously consistent with a $k=3$ cosmological coupling scenario is if \emph{some} massive stars form subsolar mass BHs that then repel the remaining infalling material, while other (presumably more massive) stars collapse directly to $10-20 M_{\odot}$ BHs  \citep[in much the same way that the progenitors of lower-mass BHs can eject material during supernova, while higher-mass BHs are thought to form via direct stellar collapse,][]{Fryer2001}.  Of course such a scenario cannot be excluded, but it would make subsolar BHs significantly more numerous than their massive counterparts \citep[based on the $P(M)\propto M^{-2.3}$ scaling of the stellar initial mass function,][]{Kroupa2001}, making their non-detection among known stellar-mass BHs (and especially the 100+ HMXBs) particularly suspect.  While the future identification of subsolar BHs in either BHXRBs or GW catalogs remains an exciting possibility \citep[e.g.,][though these would most likely be attributed to primordial BH formation in the Early Universe]{Nitz2022,TheLIGOScientificCollaboration2022}, I would argue that the current lack of theoretical and observational evidence for their existence allows us to exclude them from the current analysis.}

\subsection{BH and GC Co-Evolution}

A second fundamental assumption of this work has been that the BHs formed at the same time as the rest of the stellar population of NGC 3201.  While star formation in most GCs terminates soon after cluster formation, there still exists the possibility that BHs could be formed at late times through the mergers of binary NSs.  However, detailed modeling of GC evolution has shown that dynamically-formed NS mergers only occur in GCs that have ejected most of their BHs through three-body encounters \citep{ye2020}, since the presence of a significant BH subsystem pushes NSs out of the dense central regions where binaries can be dynamically formed.  {The same argument is true for the possibility of NS-star collisions and tidal disruptions, which, while they may occur frequently in core collapsed and massive clusters, are negligable in non-core collapsed smaller clusters like NGC 3201 \citep[][]{Kremer2022}}\footnote{Model 3 from Table 1 of \citet{Kremer2022} (which produces zero NS-star collisions) has nearly identical present-day mass and radii to the best-fit dynamical model of NGC 3201 from \citet[][Model 5]{Kremer2018}}.  Both the presence of the three RV BHs and comparisons to Monte Carlo $N$-body simulations \citep{Askar2018,Weatherford2020}, including bespoke models of NGC 3201 \citep{Kremer2018}, have suggested the presence of 10s to 100s of BHs in the cluster.  We can also safely ignore the presence of binary NS mergers that may have arisen from massive binary stars present at cluster formation, since these would have merged within a few 100 Myrs of cluster formation \citep[e.g.,][]{Chruslinska2018}, within the uncertainties in NGC 3201's age.

{The aforementioned GC simulations assumed BHs that formed with masses greater than $2.5M_{\odot}$ following the Fryer et al. 2012 prescriptions.  If BHs were instead formed at masses closer to $\sim 0.1M_{\odot}$, two-body relaxation would naturally push these low-mass particles to the outskirts of the cluster as more massive objects (e.g. NSs and massive stars) segregate into the central regions \citep[e.g.,][]{binney2008}.  For NSs, this would occur within $\sim (\left<M\right>/M_{\rm NS}) T_{\rm rel}$, where $\left<M\right>$ is the average mass of stars in the cluster, and $T_{\rm rel}$ is the half-mass relaxation time of the cluster \citep[e.g.,][]{spitzer1987}.  Assuming an average mass of $0.6M_{\odot}$ (from a zero-age Kroupa IMF), neutron star masses of $1.4M_{\odot}$, and an initial relaxation time of approximately 2 Gyr at cluster formation \citep[using the initial conditions of the best-fit NGC 3201 model from][]{Kremer2019} this yields yields a mass segregation timescale of about 1 Gyr.}

{In other words, in a cluster with $0.1M_{\odot}$ cosmologically-coupled BHs, the initial phase of cluster core collapse would be largely driven by neutron stars and other massive objects (including any BHs that \emph{did} form at larger masses), though at a significantly later time than the BH-driven collapse typically considered in our models.  However, as the BHs grew over time (and the massive stars in the cluster continued to die), the cluster would experience a kind of mass inversion at late times, where the BHs would overtake the NSs in mass, migrate back into the center, and form a more classically core collapsed cluster (until BH interactions ejected the majority of the BHs).  This also has interesting implications, since our current understanding is that ``classically'' core collapsed clusters (that is, those whose surface brightness profiles steadily increase to the center, as measured by observers) have dynamically ejected all their BHs by the present day \citep{Breen2013,Kremer2019}.  To explain core collapsed clusters under this framework would require either an extremely rapid dynamical ejection of BHs from the core (though extremely compact initial conditions) or an alternate mechanism for removing BHs from the cluster (e.g., removal of the $0.1M_{\odot}$ BHs from the outer regions of the cluster by galactic tidal fields).  }

{Secondly, if all BHs in GCs were $k=3$ BHs, then the BHs that were born with HMXB-like masses should also continue to grow redshift.  Assuming GC formation redshifts of $z=2-6$ \citep[e.g.,][]{El-Badry2018} and a population of $20M_\odot$ BHs at formation, most GCs should now host intermediate-mass BHs (IMBHs) with masses in the $500-7000M_{\odot}$ range.  These extremely-massive BHs would have dynamically decoupled from the rest of the cluster early on \citep[via the so-called ``Spitzer Instability'',][]{Spitzer1969}, and would rapidly eject one another through three-body encounters, leaving at most one IMBH in each cluster by $z=0$ \citep[e.g.][]{Kulkarni1993}.  However, there remain no secure detections of IMBHs in GCs \citep{Greene2020}, while radio observations of galactic GCs have placed strong constraints on IMBHs of more than $2000M_{\odot}$ in almost all surveyed GCs \citep[][Figure 2]{Tremou2018}.}

\section{Discussion and Conclusions}
In this paper, I have argued that the mass function of the two (or potentially three) BHs in NGC 3201 can be used to place strong constraints on the cosmological coupling between BHs and an expanding Universe.  By using the nominal age of the cluster (11.5 Gyr), I showed that the necessary growth factor of $(1+z)^3\approx55$ from $z=2.8$ to the present day is incompatible with the low masses of the two (or possibly three) BHs in NGC 3201 and our understanding of the minimum BH mass.  {For all BHs to be the $k=3$ horizonless BHs considered by \citetalias{Farrah2023a}, either both NGC 3201 BHs must be nearly face on to Earth, or BHs must form from stellar collapse with subsolar masses.}


I have focused so far only on the two (or possibly three) BHs in NGC 3201, since the small masses and old GC age make them ideal tests of cosmological BH growth.  This ignores the potential limits that could be imposed by including additional BHs with known masses from the galactic field, such as the two detached BH-star binaries identified in Gaia DR3 by \cite{El-Badry2023,El-Badry2023a}.  Unlike the two NGC 3201 BHs, Gaia BH1 and BH2 have measured masses with upper bounds (at $9.62\pm0.18M_{\odot}$ and $8.9\pm0.3M_{\odot}$, respectively) derived from the combination of RV and astrometric observations.  However, the measured ages of the BH companions are significantly less constrained (at $\gtrsim 1$ Gyr and between 5-13 Gyr, respectively) than NGC 3201's stellar population, which complicate our ability to use them as probes of BH formation at high redshift \citep[though the chemical abundances of Gaia BH2's red giant companion suggest it likely formed in the thick galactic disk, implying an age between $\sim 8-12$ Gyr; see][Section 3.8]{El-Badry2023a}.   

The analysis here is sufficient to strongly disfavor the $k=3$ coupling required for BHs to be the primary driver of $\Omega_\Lambda$.  However, placing more reliable limits on lower coupling constants, especially $k\sim 0$, would require a more sophisticated statistical treatment of both the measurement uncertainties in the RV curves and in the age determination (e.g.~using the Bayesian age fitting of \citealt{wagner-kaiser2017}).  The mass distributions can also be improved, since the uninformative prior I have employed (flat, except for the existence of the mass gap) can be replaced with a prior on BH mass distribution \citep[e.g.][]{Ho2011} near the NS/BH boundary taken from LIGO/Virgo observations (which will only improve after the upcoming O4 observation run).   Finally, constraining the inclination of the RV BHs would allow us to place firm upper limits on the BHs mass, which may be possible with diffraction-limited telescopes (e.g., HST, VLT) or with JWST astrometry observations, either of which may be capable of resolving the 0.2 mas orbital motion of BH1 \citep{Giesers2018}.

\begin{acknowledgements}

I am grateful to the anonymous referee for several useful suggestions, as well as Michael Zevin, Duncan Farrah, Kevin Croker, Greg Tarle, Kurtis Nishimura, Sara Petty, Adrienne Erickcek, Andrew Mann, and Kareem El-Badry for useful comments.  CR acknowledges support from the Alfred P.~Sloan Foundation and the David and Lucile Packard Foundation.

\end{acknowledgements}
\bibliography{sample631}

\end{document}